\documentclass[
reprint,
superscriptaddress,
amsmath,amssymb,
prl,
twocolumn,
floatfix
]{revtex4-1}

\usepackage{amsthm}
\usepackage{graphicx}
\usepackage{dcolumn}
\usepackage{bm}
\usepackage{mathtools}
\usepackage{color}

\begin{document}
	\title{Bifurcation without parameters in a chaotic system with a memristive element}
	\author{Tom Birkoben}
	\email{tobi@tf.uni-kiel.de}
	\affiliation{Nanoelektronik, Technische Fakult\"at, Christian-Albrechts-Universit\"at zu Kiel, Kaisertra{\ss}e 2, 24143 Kiel, Germany}
	\author{Moritz Drangmeister}
	\email{drangmeister.m@gmail.com}
	\affiliation{Institut f\"ur Theoretische Physik, Technische Universit\"at Berlin, Hardenbergstra{\ss}e 36, 10623 Berlin, Germany}%
	\author{Finn Zahari}
	\email{fnz@tf.uni-kiel.de}
	\affiliation{Nanoelektronik, Technische Fakult\"at, Christian-Albrechts-Universit\"at zu Kiel, Kaisertra{\ss}e 2, 24143 Kiel, Germany}
	\author{Serhiy~Yanchuk}
	\email{yanchuk@math.tu-berlin.de}
	\affiliation{Institut f\"ur Mathematik, Technische Universit\"at Berlin, Stra{\ss}e des 17. Juni 136, 10623 Berlin, Germany}%
	\author{Philipp H\"ovel}
	\email{philipp.hoevel@ucc.ie}
	\affiliation{School of Mathematical Sciences, University College Cork, Western Road, Cork T12 XF64, Ireland}
	\affiliation{Institut f\"ur Theoretische Physik, Technische Universit\"at Berlin, Hardenbergstra{\ss}e 36, 10623 Berlin, Germany}%
	\author{Hermann Kohlstedt}
	\email{hko@tf.uni-kiel.de}
	\affiliation{Nanoelektronik, Technische Fakult\"at, Christian-Albrechts-Universit\"at zu Kiel, Kaisertra{\ss}e 2, 24143 Kiel, Germany}
	\date{\today}
	\begin{abstract}
	We investigate the effect of memory on a chaotic system experimentally and theoretically. For this purpose, we use Chua's oscillator as an electrical model system showing chaotic dynamics extended by a memory element in form of a double-barrier memristive device. The device consists of Au/NbO$_\text{x}$/Al$_\text{2}$O$_\text{3}$/Al/Nb layers and exhibits strong analog-type resistive changes depending on the history of the charge flow. In the extended system strong changes in the dynamics of chaotic oscillations are observable. The otherwise fluctuating amplitudes of the Chua system are disrupted by transient silent states. After developing a model for Chua's oscillator with a memristive device, the numerical treatment reveals the underling dynamics as driven by the slow-fast dynamics of the memory element. Furthermore, the stabilizing and destabilizing dynamic bifurcations are identified that are passed by the system during its chaotic behavior.
	\end{abstract}
	\pacs{Valid PACS appear here}
	\maketitle
\begin{figure}[ht!]
    \centering
    \includegraphics[width=\linewidth]{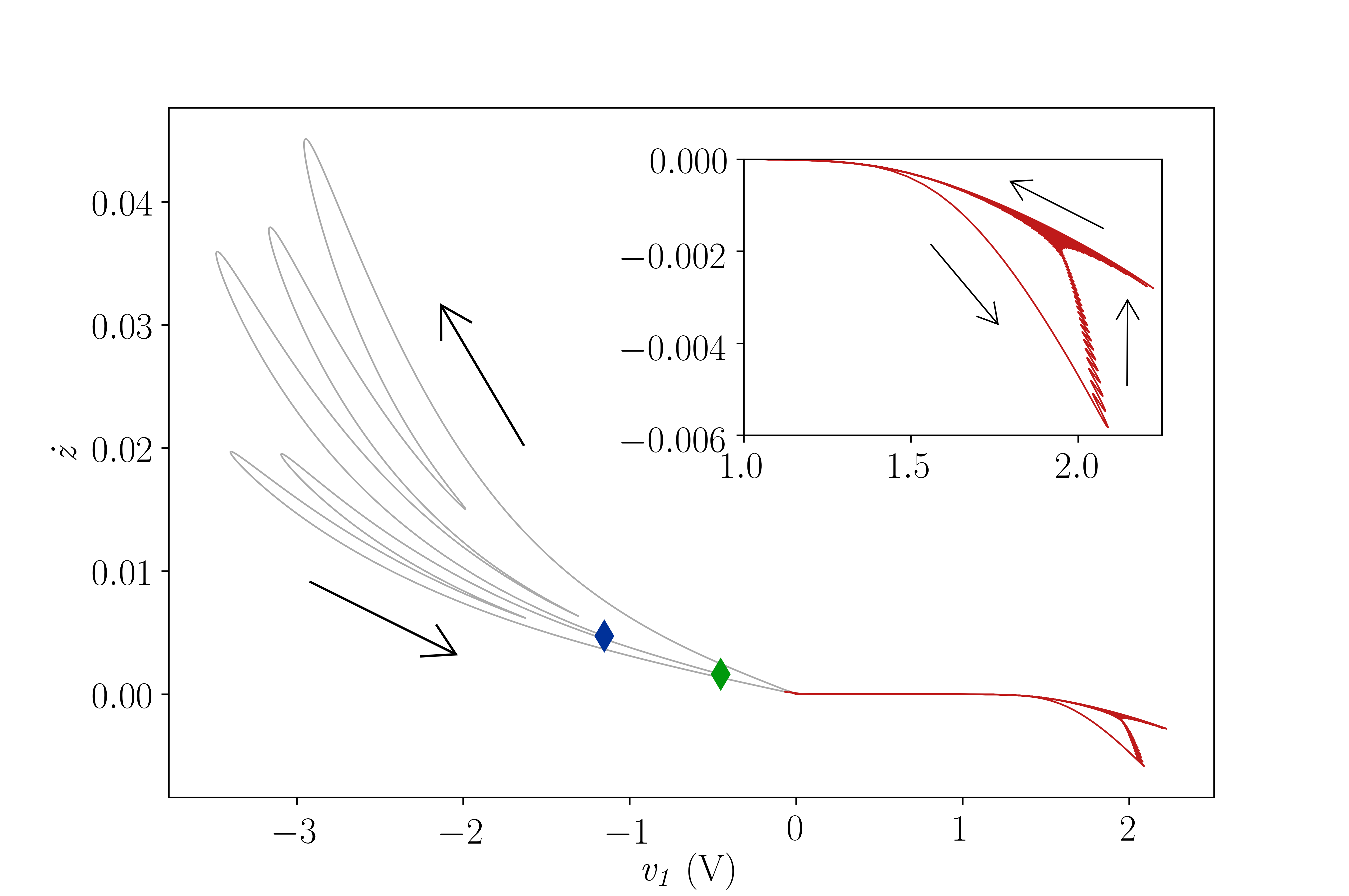}
    \caption{The gradient of $z$ as a function of the state variable $v_1$ along the trajectory. The red trace corresponds to the slow decrease of $z$ during the transient silent state event. The grey part refers to the setback. The green and blue diamond marks mark the start and the endpoint, respectively.}
    \label{fig:grad_z}
\end{figure}

The field of nonlinear dynamics, chaos, and complexity has attracted increasing interest from the point of fundamental science and engineering during the last decades \cite{schuster_deterministic_2005, strogatz_nonlinear_2015}. Classical and well-explored nonlinear phenomena form a fundamental scientific repertoire to shed more light on novel interdisciplinary research areas such as complex network systems. To name but a few, this includes spatio-temporal pattern formation in chemical reactions, pulse coupled oscillators, chaotic weather formation or time-delay systems \cite{epstein_nonlinear_1996,pikovsky_synchronization:_2003,Yanchuk2017}. \newline
Currently, time-varying networks, neuroscience, and social dynamics are areas of intense research efforts in nonlinear science \cite{sporns_networks_2011,osipov_synchronization_2007,skarda_how_1987,buzsaki_rhythms_2006}. Furthermore, nonlinear systems with experimentally observable chaotic signatures have received much attention, as they are widely distributed over many different fields including optical, mechanical and chemical systems \cite{epstein_nonlinear_1996,gilet_chaotic_2009,ievlev_intermittency_2014,brzeski_analysis_2015,buzsaki_rhythms_2006}. In this context, electronic systems are of particular interest. Rather simple analog circuits allow the study and control of chaos and nonlinear dynamical phenomena. The fast and easy access to system parameters in experiments through the variation of passive elements of the circuit, i.e., resistances, inductances and capacitances, is an effective way to tune the circuit dynamics and to observe the results in real time. The first chaotic circuit was realized by Leon Chua in the 1980s, consisting of three energy storing elements and a nonlinear electronic device. The circuit exhibits a classical period-doubling route to chaos as well as a chaotic double-scroll attractor \cite{matsumoto_chaotic_1984,l._o._chua_universal_1993}. \newline
In this letter, we present a novel realization of Chua's circuit comprising a memristive device, i.e., a storage element \cite{vaidyanathan_advances_2017}. Memristive devices are currently investigated from the perspective of non-volatile memories and promising devices to mimic basal synaptic mechanisms in neuromorphic circuits \cite{ielmini_resistive_2016,tetzlaff_memristors_2014,kozma_advances_2012,adamatzky_networks:_2014}.In general a memristive device connects the current $I$ and voltage $V$ nonlinearly. The resistance of such a system depends on a mechanism relating the voltage to a change of an internal state $z$:
\begin{subequations}
\begin{align}\label{eq:IV_mem}
 I =& \;G(z)\cdot V, \\
 \dot{z} =& \;f(z,V).
\end{align}
\end{subequations}
In its simplest form such a device consists of a metal-insulator-metal capacitor-like structure. Here, an applied voltage can lead to the movement of ions within the insulator, resulting in a change of the resistance \cite{strukov_missing_2008}. Thus, the history of the applied voltage is connected to the current state of the device. As a result, the current-voltage characteristics or \textit{I-V} curve of a memristive device exhibits a hysteresis loop. For more details about memristive devices and the underlying physical and chemical mechanisms see {Ref.~\cite{ielmini_resistive_2016}}.
\begin{figure}[ht!]
    \centering
    \includegraphics[width=\linewidth]{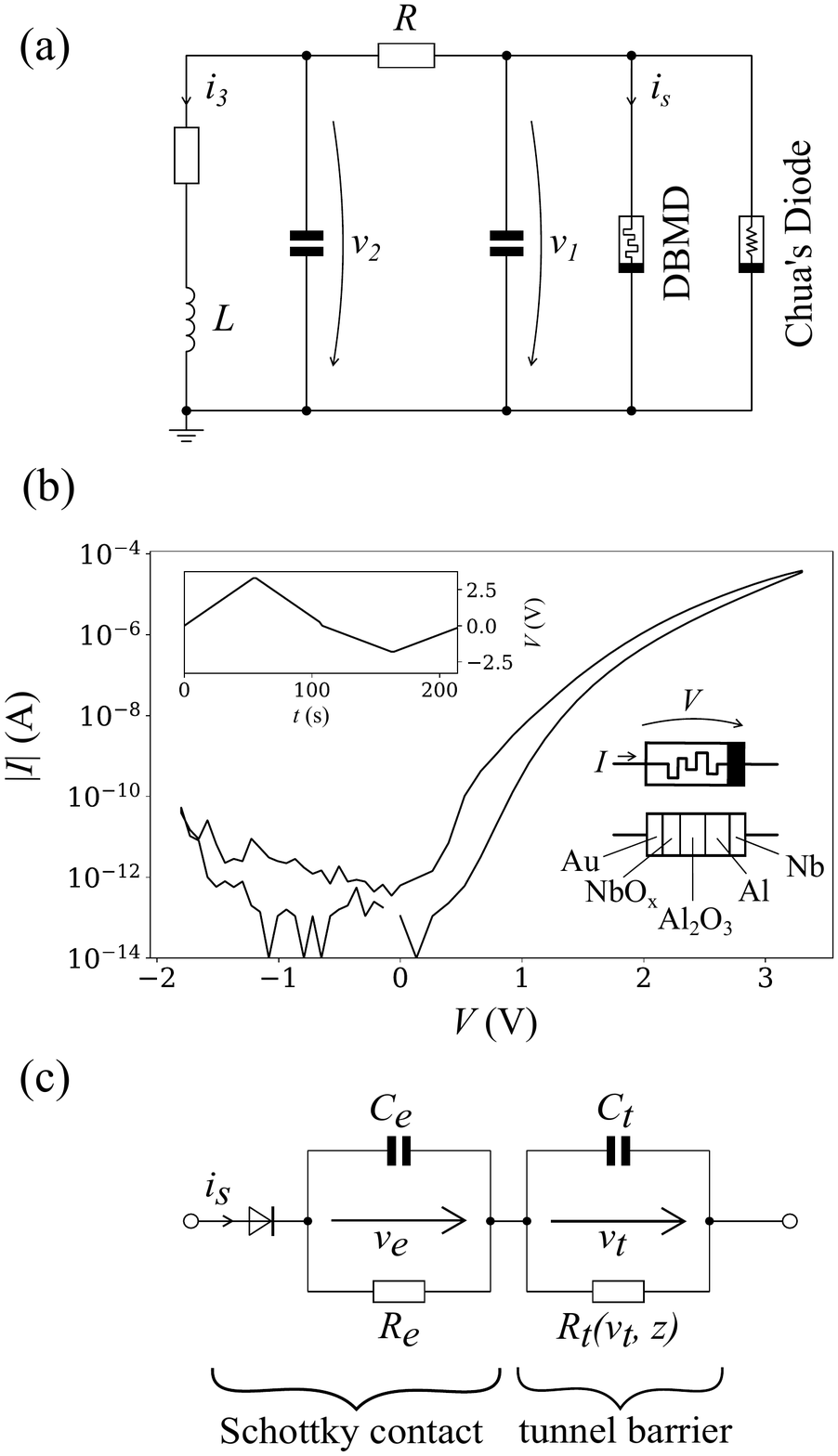}
    \caption{(a) Experimental setup of Chua's circuit comprising a double-barrier memristive device (DBMD). The device is integrated in parallel to Chua's diode. The current flow through the device discharges the capacitor resulting in an additional negative feedback on it. (b) A typical \textit{I-V} curve of a DBMD. It consists of Au/NbO$_\text{x}$/Al$_\text{2}$O$_\text{3}$/Al/Nb thin layers. As a triangular voltage (inset) is applied to the device, the current increases. This changes the internal state of the device leading to a transition from a high-resistance state to a low-resistance state. Applying a negative voltage resets the low-resistance state again to a high-resistance state. We like to emphasize that this transition is not binary but continuous. (c) Equivalent circuit of the DBMD. It can be subdivided into a Schottky contact and a tunnel barrier. The resistance changes as the applied voltage modulates the effective barrier heights of the tunnel barrier and the Schottky contact, respectively. For further details on the derivation of the model see Ref.~\cite{solan_enhanced_2017}.} 
    \label{fig:circuit}
\end{figure}
\newline
Figure~\ref{fig:circuit}(a) shows the circuit layout considered in this letter. The chaotic circuit as proposed by Leon Chua is extended with a memristive device in parallel to Chua's diode. Therefore, the solid state device superimposes the necessary nonlinearity to drive the chaotic circuit. As the state of the double-barrier memristive device (DBMD) depends strongly on the history of the applied voltages across it, the strength of the additional discharge of the capacitor varies chaotically with time. The used device consists of a Au/NbO$_\text{x}$/Al$_\text{2}$O$_\text{3}$/Al/Nb layer sequence \cite{hansen_double_2015,dirkmann_role_2016}. We emphasize that this kind of device is filament free, i.e., an interface-based switching is responsible for the pinched hysteresis observed in the \textit{I-V} curve. Furthermore, the switching is not binary but continuous as depicted in Fig.~\ref{fig:circuit} (b). The transition from a high-resistance state (HRS) to a low-resistance state (LRS) is well observable for applying a positive voltage to the Au electrode in respect to the Nb electrode. The device remains in the LRS after the polarity of the applied voltage switches, but changes its resistance again in an analog fashion to a HRS after a threshold voltage was exceeded \cite{hansen_double_2015}. Considering the internal structure of a DBMD the functional films can be modelled as an equivalent circuit as shown in Fig.~\ref{fig:circuit} (c). The metal semiconductor transition is modeled as a Schottky contact followed by a tunnel barrier. From this model the following set of differential equations, which describe the state of the DBMD, can be obtained: \cite{solan_enhanced_2017} 
\begin{subequations}
\begin{align}\label{eq:v_e-v_t-z}
\dot{v_e} =& \frac{1}{C_e} \left( i_s(v_s,z) - \frac{v_e}{R_e(z)} \right) , \\
\dot{v_t}=& \frac{1}{C_t} \left( i_s(v_s,z) - i_t(v_t,z) \right) , \\
\dot{z}=& -\frac{\hat{Z}\omega(z)}{e^{\varphi_a (v_1,z)}} \sinh\left( \frac{v_r(v_1,v_s,z)+v_e-V_c}{V_e} \right).\label{eq:z}
\end{align}
\end{subequations}
 The memory component of the DBMD is represented through the state variable $z$, which refers to the average ion-position inside the active layer, that is the NbO$_\text{x}$ solid-state electrolyte. During the switching oxygen-vacancies move and consequently decrease and increase the potential on the interface at the Schottky contact and the tunnel barrier, respectively \cite{solan_enhanced_2017}. The voltage over the Schottky contact is $v_s = v_1-v_e-v_t$ and leads to the total current $i_s(v_s,z)$ through the device as:
\begin{align}\label{eq:i_s}
    i_s(v_s,z) =& I_s \exp\left(-\varphi_s(z) - \alpha_f \sqrt{\frac{|v_s|-v_s}{\alpha_s V_\vartheta}} \right) \nonumber\\
    &\times \left[\exp\left(\frac{v_s}{n(z) V_\vartheta}\right) - 1 \right].
\end{align}
with $\varphi_s(z)$ as the state-dependent normalized Schottky-barrier height, $n(z)$ as an ideality factor and $\alpha_f$ as a fitting parameter for the Schottky-effect denoted by the normalized Schottky-barrier thickness $\alpha_s$. The amplitude of the current $I_s$ scales the total current depending on the temperature and device area, respectively \cite{solan_enhanced_2017}. \newline
Since the memristive device is implemented experimentally in parallel to Chua's diode, the voltage drop across the device is equal to the state variable $v_1$ of the original chaotic oscillator. Therefore, the current flow through the DBMD discharges the capacitor and functions as a negative feedback to the first state-variable $v_1$. The modified equations of the system augmented by the memory element are as follows:
\begin{subequations}
\begin{align}
    \dot{v_1} &= \frac{1}{C_1}\left(\frac{v_2-v_1}{R} - f(v_1) - i_s(v_s,z) \right) , \\
    \dot{v_2} &= \frac{1}{C_2}\left(\frac{v_1-v_2}{R} -  i_3 \right) , \\
    \dot{i_3} &= -\frac{1}{L} v_2 \, , \label{eq:v_1-v_2-i_3} 
\end{align}
\end{subequations}
with Chua's diode modelled as a piece-wise linear function:
\begin{align}\label{eq:chua}
    f(v_1) = m_0 v_1 + \frac{m_1-m_0}{2}\left(|v_1+B_p|-|v_1-B_p| \right).
\end{align}
The closed system (\ref{eq:v_e-v_t-z})-(\ref{eq:v_1-v_2-i_3}) describes the dynamics of the complete circuit augmented with a DBMD (additional information in~\cite{SI}).
\begin{figure*}%
\includegraphics[width=\columnwidth]{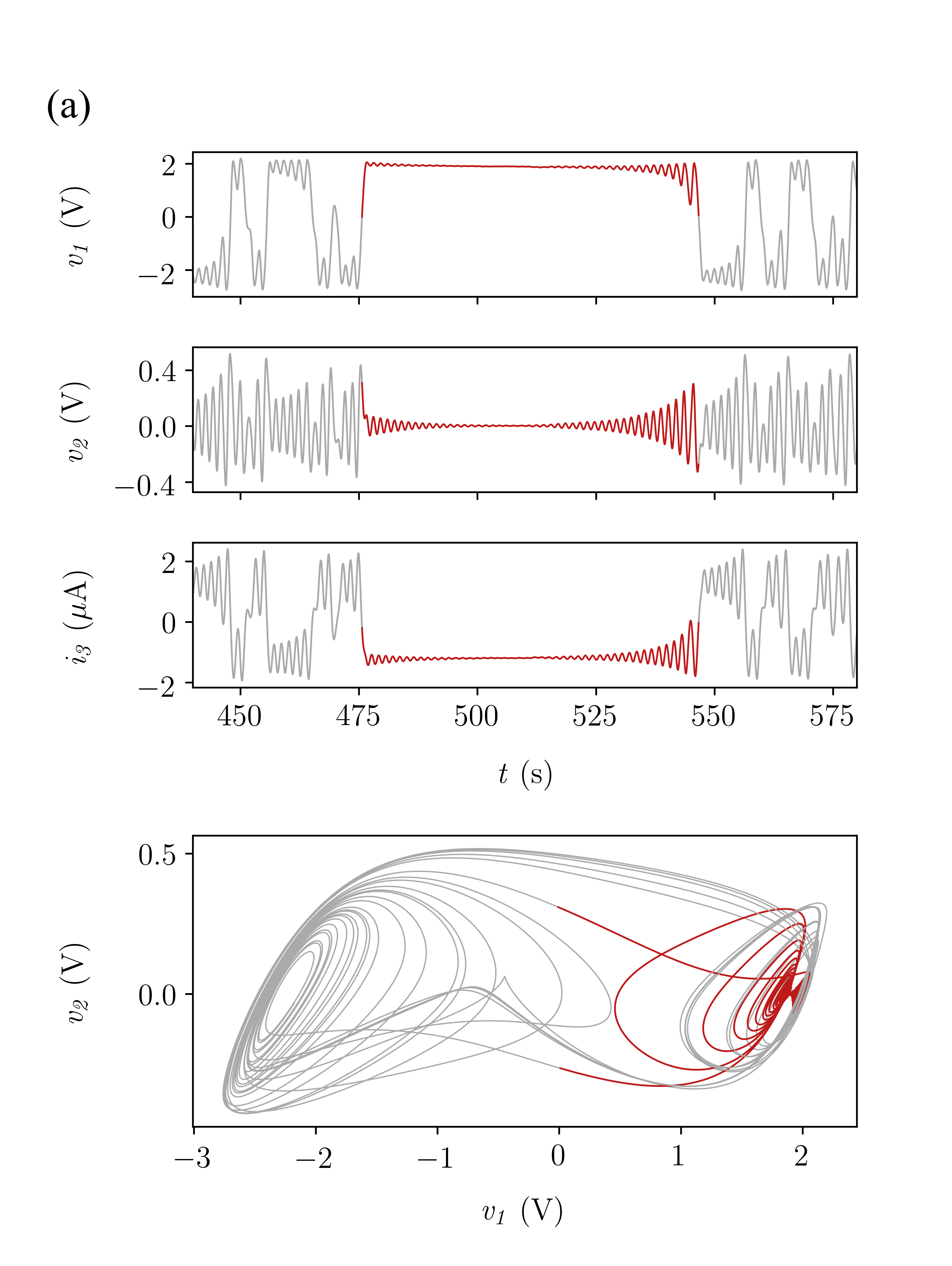}%
\includegraphics[width=\columnwidth]{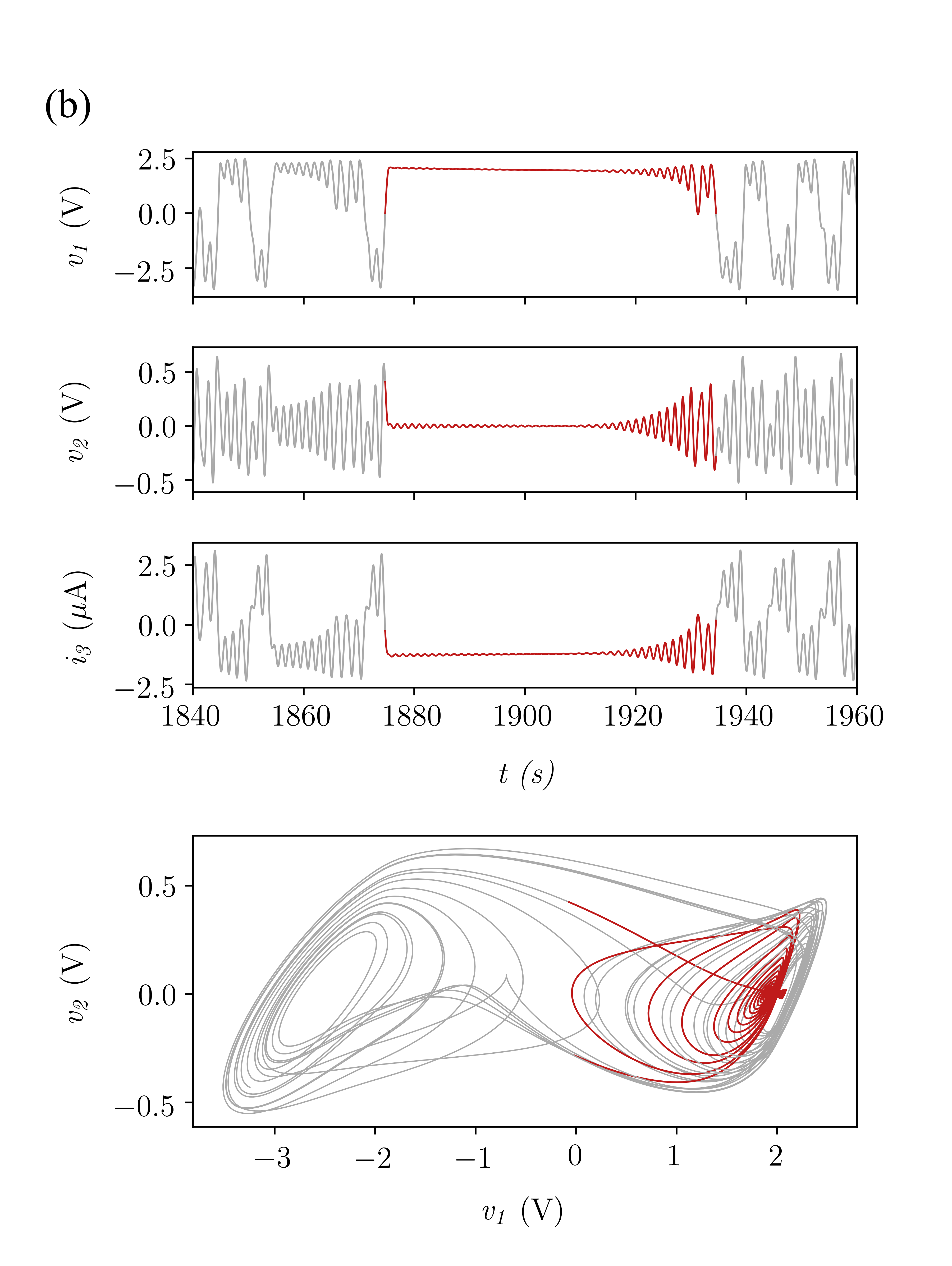}%
\caption{(a) Snapshot of the experimental time series measurement of Chua's circuit comprising a DBMD. The top three traces show the time evolution of the fundamental state variables of the system. The chaotic oscillations (grey) are disrupted by minute long dampening of the intrinsic oscillations (red). The trajectory of the system will leave this transient silent state and return to the well known chaotic oscillations. The lower diagram shows the $(v_1,v_2)$ phase space. (b) Snapshot of the numerical results of the extented chaotic system comprising a DBMD. Damping and concurrent excitation of the Chua variables $v_1$, $v_2$ and $i_3$ can be observed as well and are in good agreement to the experimental results (additional information in~\cite{SI}).}%
\label{fig:tss_exp_sim}
\end{figure*}
\newline 
As one can observe from Fig.~\ref{fig:tss_exp_sim}, the additional memory element has a significant influence on the system dynamics. In the original system without the  memristive device, the local oscillations are amplified until the trajectory switches to the opposite side of the characteristic double-scroll attractor. With the introduction of the memristive device the purely chaotic dynamics are interrupted by transient silent states (TSS). These states are characterized by the damping and relatively long time-intervals of almost constant voltages and currents. After a period nearly without any oscillations, an onset of the local oscillations follows. The strong diode-like characteristic of the DBMD diminishes the influence of the it on the other side of the chaotic double-scroll attractor. This results in the clearly asymmetric change of the system behavior.
\begin{figure}[pt!]
    \centering
    \includegraphics[width=\linewidth]{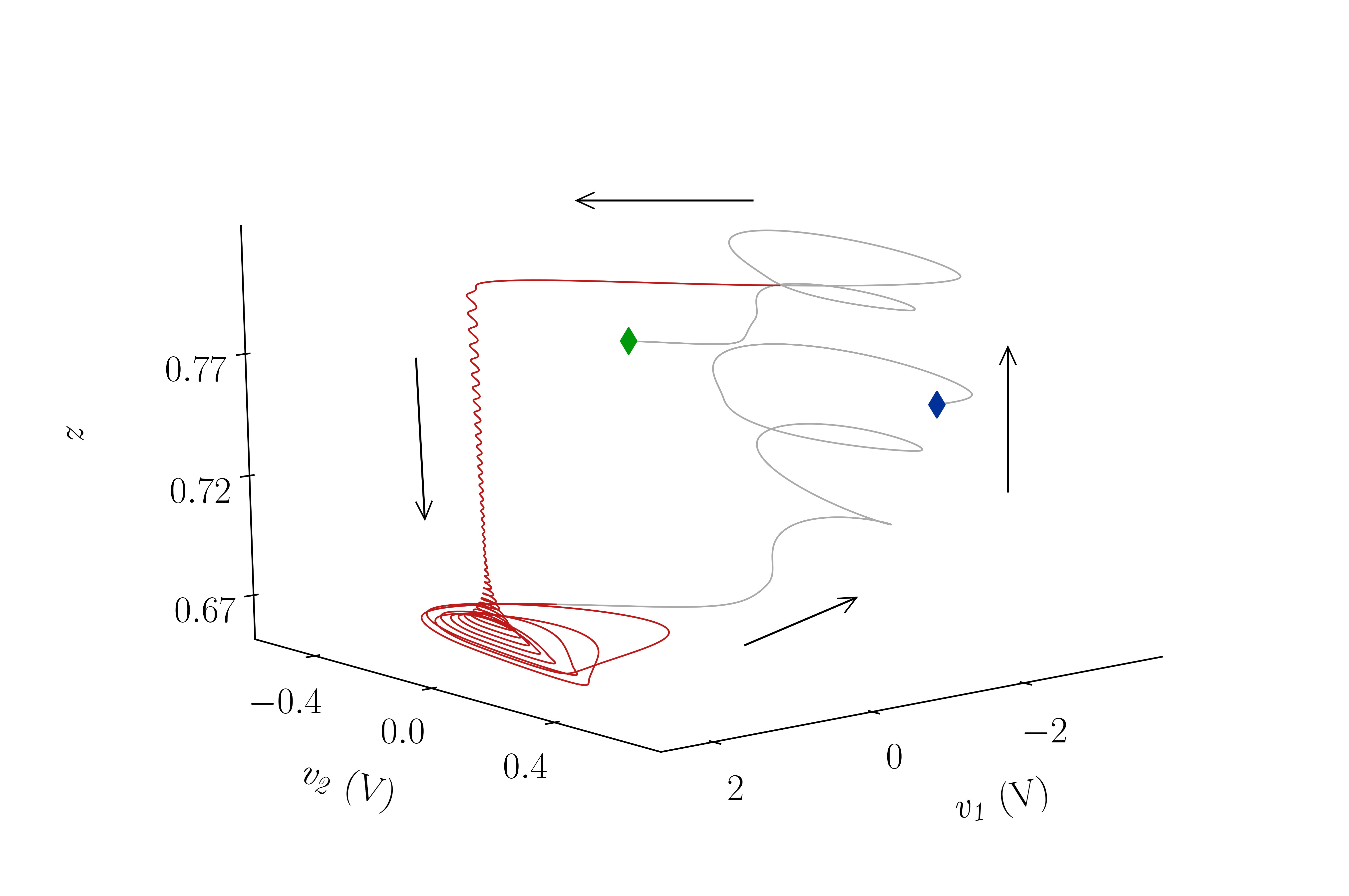}
    \caption{Trajectory of the chaotic system, exhibiting a transient silent state by lowering the resistance and the successive setback to a high-resistance state. The evolution starts at the green diamond and stops at the blue one. The decrease of $z$ is related to the onset of the dumping of the oscillations.}
    \label{fig:one_circle_TSS}
\end{figure}
\newline
The numerical solution shows clearly the role of the internal state variable of the DBMD. In Fig.~\ref{fig:one_circle_TSS} the time series for one typical TSS is shown in the $(v_1,v_2,z)$ phase space. The decrease of the internal state $z$ can be observed over time, which  leads to a higher conductance of the device and a damping of the oscillations. It is followed by a steep increase when the resuming chaotic oscillations lead to a negative voltage over the memristive device. 
\begin{figure}[ht!]
    \centering
    \includegraphics[width=\linewidth]{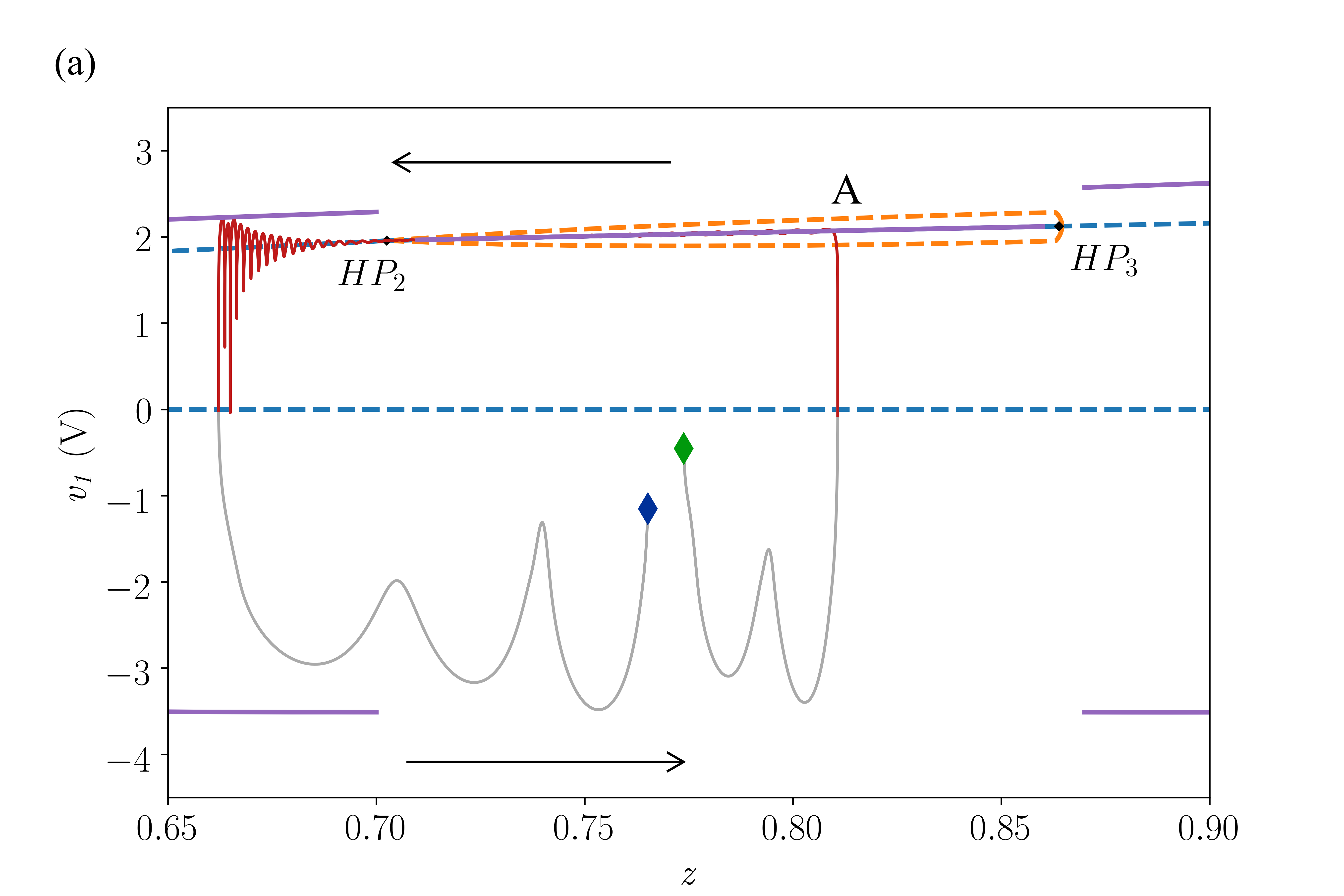}
    \includegraphics[width=\linewidth]{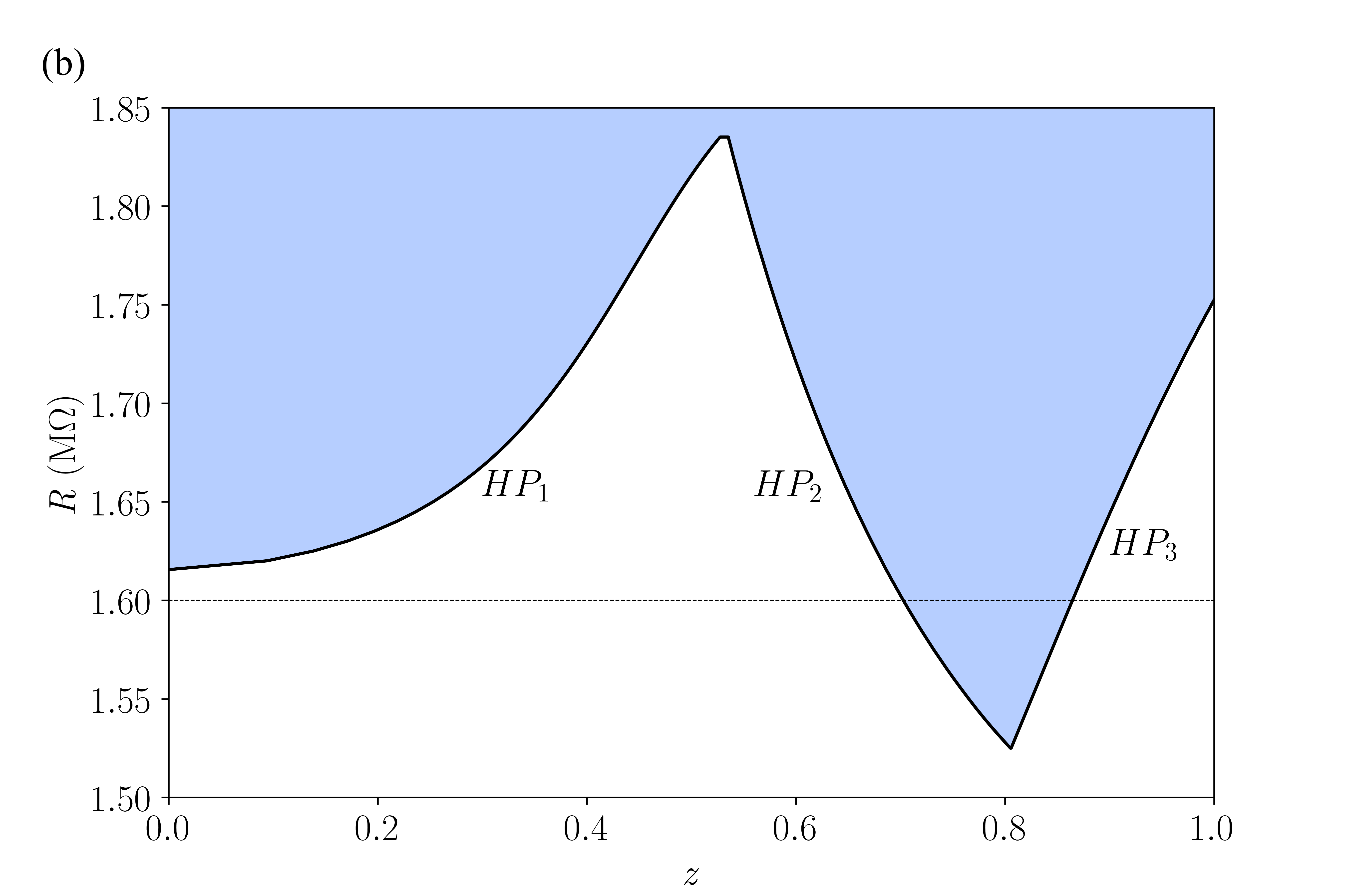}
        \caption{(a) Bifurcation diagram for $v_1$ over control parameter $z$ with stationary solutions in blue, periodic solutions in orange and chaotic solutions in purple. Continuous lines depict stable solutions, dashed lines unstable ones. For periodic and chaotic attractors, the minimum and maximum of $v_1$ are shown. Superimposed to the bifurcation diagram is a transient silent state with a green diamond and blue diamond corresponding to the starting and stopping point, respectively. The trajectory follows the arrows. (b) Position of the Hopf bifurcation points (HP) parametrized by $R$ if $z$ is varied as a control parameter. The blue shaded area shows where a stable branch exists in the solution. During the experiment $R$ is fixed at 1.6 M$\Omega$ (dashed, horizontal line).}
    \label{fig:bifurcation_diagram}
\end{figure}
\newline 
In the following we show that the chaotic dynamics with TSS events can be understood as a slow-fast motion with the slowest timescale governed by the memory $z$. More specifically, the variable $z$ can be considered as a control parameter for the remaining faster variables, i.e., $\dot{z} = 0$ \cite{Kuehn2015}. For the experimentally chosen value $R=1.6$ M$\Omega$, the fast dynamics possess a stable stationary state $(v_e(z),v_t(z),v_1(z),v_2=0, i_3(z))$ for all values of $z$ in the interval $0.71<z<0.86$, see solid blue line in Fig.~\ref{fig:bifurcation_diagram}(a). At the boundaries, $z=0.71$ and $z=0.86$, the branch becomes unstable in subcritical Hopf bifurcations. This set (solid blue line in Fig.~\ref{fig:bifurcation_diagram}(a)) is called the stable slow manifold \cite{Kuehn2015}, and the trajectories of the full system are attracted to this manifold on the fast timescale. Being close to the manifold, the dynamics are then governed by the single scalar equation~(\ref{eq:z}) for the $z$ variable with all fast variables being confined to the manifold. From the differential equation~(\ref{eq:z}) of the internal state $z$, it can be seen that $z$ decreases on the slow manifold as positive values for $v_1$, $v_e$, and $v_t$ lead to a negative sign of the derivative. This corresponds to motion along the manifold to the left. The above multi-scale arguments explain parts of the dynamics observed in the full system and are shown in Fig.~\ref{fig:bifurcation_diagram}(a). Once the trajectory is attracted to the slow manifold (point A in Fig.~\ref{fig:bifurcation_diagram}(a)), the TSS episode starts and the memory variable $z$  decreases slowly until it reaches the Hopf point, $HP_2$, at $z = 0.71$. At this point, the slow manifold becomes unstable and the trajectory exhibits amplified oscillations, while $z$ still decreases further. When the oscillations become sufficiently large, the orbit leaves the neighborhood of the slow manifold and the TSS episode ends. After the TSS has terminated, the voltages  $v_1$, $v_e$, and $v_t$ decrease and the variable $z$ accelerates and approaches the fast timescale, see Fig.~\ref{fig:grad_z} for the gradient $\dot{z}$ along the trajectory. In this way, a slow setting to a LRS and a faster reset to a HRS drives the slow fast dynamics behind the TSS. Figure~\ref{fig:bifurcation_diagram}(b) shows the changing size of the stable part of the slow manifold, this is the lengths of the TSSs phases, depending on $R$.  In particular, the stable branch can be seen to expand with increasing $R$, until the stability becomes independent of $z$ when $HP_1$ and $HP_2$ meet. For decreasing $R$, the stable branch shrinks and vanishes with $HP_2$ and $HP_3$.
\newline 
In summary we have described the influence of a new solid state memory device on a model system exhibiting chaotic dynamics to study the influence of memory on chaos. Interestingly, the intrinsic memory of the DBMD has a stabilizing and ordering effect on the otherwise purely chaotic motions of the system. The experimental observation as well as the subsequent theoretical treatment reveal the underlying dynamics. We have observed that the memory acts on the slowest timescale, and thus, it can be considered as an intrinsic slowly-changing bifurcation parameter. In particular, when the fast chaotic motion of the system is interrupted by a TSS, the dynamics can be considered as a steady state, which is adiabatically changing with the slow variation of the memory. Then, the chaotic oscillations are dumped until the stability changes. Such a change occurs when the memory reaches a threshold, and a Hopf bifurcation of the fast system leads to amplification of chaotic oscillations. \newline
As the memristive device in use has two time scales for the set and setback of the resistive states, this dumping and successive amplification is driven by these slow-fast dynamics. Although this study is restricted to an electronic system, the observed behavior and general dynamical changes might be found in a broader range of systems. The occurrence of transiently stable and ordered behavior in otherwise highly nonlinear or chaotic trajectories might be related to constant dynamic bifurcations as an intrinsic memory element of these systems changes its state. The sequence of these transient states is primary driven by the underlying chaotic motions but the duration of and recovery from each depends strongly on the characteristic timescales of this memory element. Speaking in more general terms it is driven by the time needed to adapt to a new input. \newline
\begin{acknowledgements}
The authors acknowledge financial support by Deutsche Forschungsgemeinschaft (DFG, German Research Foundation), projects RU2093 (TB, FZ, HK), 411803875 (SY) and Collaborative Research Center 910 (MD, PH).
\end{acknowledgements}

\bibliographystyle{apsrev4-1}

%

\end{document}